\documentclass[a4paper]{cas-sc}

\usepackage{natbib}
\usepackage{amsfonts}
\usepackage{amsmath}
\usepackage{threeparttable}
\usepackage{tablefootnote}
\usepackage{multirow}
\usepackage{subcaption}

\makeatletter
\newcommand{\rmnum}[1]{\@roman{#1}}
\newcommand{\Rmnum}[1]{\@Roman{#1}}
\makeatother

\def\tsc#1{\csdef{#1}{\textsc{\lowercase{#1}}\xspace}}
\tsc{WGM}
\tsc{QE}
\tsc{EP}
\tsc{PMS}
\tsc{BEC}
\tsc{DE}
\begin{document}
\let\WriteBookmarks\relax
\def\floatpagepagefraction{1}
\def\textpagefraction{.001}
\shorttitle{Bayesian vector autoregressive analysis}
\shortauthors{Jieling Jin}
%\begin{frontmatter}

\title [mode = title]{Bayesian vector autoregressive analysis of macroeconomic and transport influences on urban traffic accidents}                      

%\tnotetext[1]{This document is the results of the research project funded by the National Science Foundation.}

\author{Jieling Jin}[orcid=0000-0002-2063-5156]
\ead{jielingkim@csu.edu.cn}

\credit{Conceptualization of this study, Methodology, Software, Original draft preparation}

\address{Urban Transport Research Center, School of Traffic and Transportation Engineering, Central South University, Changsha, Hunan, China}

\begin{abstract}
The macro influencing factors analysis of urban traffic safety is important to guide the direction of urban development to reduce the frequency of traffic accidents. In this study, a Bayesian vector autoregressive(BVAR) model was developed to exploring the impact of six macro-level economic and transport factors, including population, GDP, private vehicle ownership, bus ownership, subway rail mileage and road average speed on traffic accidents with the small sample size transport annual report data in Beijing. The results show that the BVAR model was suitable for time series analysis of traffic accidents in small sample situations. In macroeconomic factors, GDP growth was considered to reduce the number of traffic accidents in the long term, while population growth had a positive effect on traffic accidents in the short term. With the respect to macro-transport factors, road average speed and private vehicle ownership was perceived to increase traffic accidents in long duration, whereas bus ownership and subway rail mileage had long-term negative effects, with the greatest positive effect for road average speed and the greatest negative effect for subway rail mileage. This study suggests that government departments can reduce the number of traffic accidents by increasing investment in public transportation infrastructures, limiting private vehicles and road speed.
\end{abstract}

\begin{highlights}
\item Bayesian vector autoregressive model is introduced to solve the problem of time series analysis in the case of small samples for traffic accident analysis.
\item Identified that urban economic growth contributes to the reduction of traffic accidents in the long term, while population growth has a positive effect on traffic accidents in the short term.
\item Revealed that road average speed and private vehicle ownership is considered to increase traffic accidents in long duration, and the positive effect is greatest for road average speed.
\item Uncovered that bus ownership and subway rail mileage have a long-term negative effect, and the negative effect is greatest for subway rail mileage.

\end{highlights}


\maketitle

\section{Introduction}

According to the \cite{WHO2021} statistics, approximately 1.3 million people die each year as a result of road traffic crashes, between 20 and 50 million more people suffer non-fatal injuries, with many incurring a disability as a result of their injury, and road traffic crashes cost most countries $3\%$ of their gross domestic product.  Traffic safety has become one of the focus issues in the field of transportation, due to the serious hazards of road traffic accidents. Exploring the influencing factors of traffic accidents has attracted the attention of many researchers. Many researchers have found that four categories of micro factors in the elements of the transport system, including human\citep{zhang2019human}, vehicle\citep{almeida2013man}, road\citep{wu2020economic} and environment\citep{lankarani2014impact} significantly impact traffic accidents. 
And with recent advances in computer technologies and sensing technologies for micro factors, scholars have made some achievements in analysing the micro influences on traffic accidents which have already contributed to the development of intelligent transportation technologies such as vehicle-road collaboration\citep{qu2019curve} and driverlessness\citep{gang2019safety}.

Researchers have also begun to discuss how urban macro-level factors can impact road traffic accidents. The economic developments have been shown to have some positive\citep{lee2008analysis} or negative\citep{apparao2013identification} impact on the number of traffic accidents\citep{sirajudeen2021sources}, and other macroeconomic factors also be identified some relationship between them and traffic accidents, such as there is a positive contribution of population growth to the number of road accidents\citep{li2018exploring}. Moreover, the number of traffic accidents is affected by macro-transport factors such as private vehicle ownership\citep{sun2019road} and road average speed\citep{wang2009effects} have a positive contribution to traffic accidents, while public transport infrastructure has a negative impact\citep{soehodho2017public}. However, the studies that simultaneously analyze the temporal dynamic relationship between traffic accidents and macroeconomic and transport factors continue to be lacking.

Most of the existing studies on the analysis of traffic accident influencing factors have focused on statistical methods. The linear regression\citep{iwata2010relationship}, logistic regression\citep{liu2020impact}, structural equation models\citep{najaf2018city} and other statistical models\citep{ghasedi2021prediction} are used to explain the the correlation between accidents and road alignment characteristics, economic growth and other influences. In addition, while these traditional statistical analysis methods can analysis the influencing factors of accidents, they are difficult to reveal the dynamic relationship among the factors and accidents, since they do not consider temporal characteristics. 

Some time-series analysis methods have also be explored to analysis traffic accident influences in previous studies. For instance, a combination of autoregressive distributed lag and vector error correction model is used to determine the short or long term causal relationships between the number of road accidents and socio-economic development\cite{li2018exploring},  a vector autoregressive model is developed to explore the dynamic relationship between motorway collisions and road infrastructure, social demographics, traffic and weather characteristics\cite{michalaki2016time}.
These time series models can find the temporal characteristics of accident influencing factors, but the traffic accident multivariate time series analysis in the case of small samples is still a difficult task.

The urban traffic accidents time-series data which containing macro-level economic and transport variables are mostly collected by annual city statistics report, and with multivariate and small sample characteristics. We proposed to use a Bayesian vector autoregressive(BVAR) model for analyzing the impact of the macroeconomic and transport factors on traffic accidents based on the annual city statistics time series data. The BVAR model has been widely used by researchers in the field of macroeconomic\citep{ma2021application,giannone2014short,dejong2000bayesian} due to its advantages of being able to deal with time-series data and achieving good parameter estimation results with small samples, and the BVAR study in the field of traffic safety is beginning to appear\citep{li2019bayesian}.

This paper mainly solves the following two problems:
(1) Determining the Bayesian vector autoregressive model can be used to obtain explore the small sample size time series data structure in the traffic accident analysis field;
(2) Exploring the dynamic influence of the macro-level economic and transport factors  on traffic accidents.

The rest of the paper is organized as follows: 
The data sources and preparation are given in section 2. 
Section 3 describes the methodology adopted. Section 4 introduces the detailed results and discussion. 
Section 5 gives conclusions and directions for future work.

\section{Data preparation}

\subsection{Data collection}

In this study, the dataset was collected from the Beijing Transport Development Annual Report from 2003 to 2020 published by the Beijing Transport Institute. This study covers all 13 administrative regions of Beijing. 
According to relevant studies and data characteristics of the annual reports, seven time series variables including the number of annual traffic accidents were selected for analysis. 
Other variables in the dataset could be divided into two types: macroeconomic and transport factors, macroeconomic variables included urban GDP and population, while macro-transport variables included urban subway rail mileage, bus ownership, private vehicle ownership and road average  speed.

\begin{figure}[h!]
	\centering
	\includegraphics[scale=0.85]{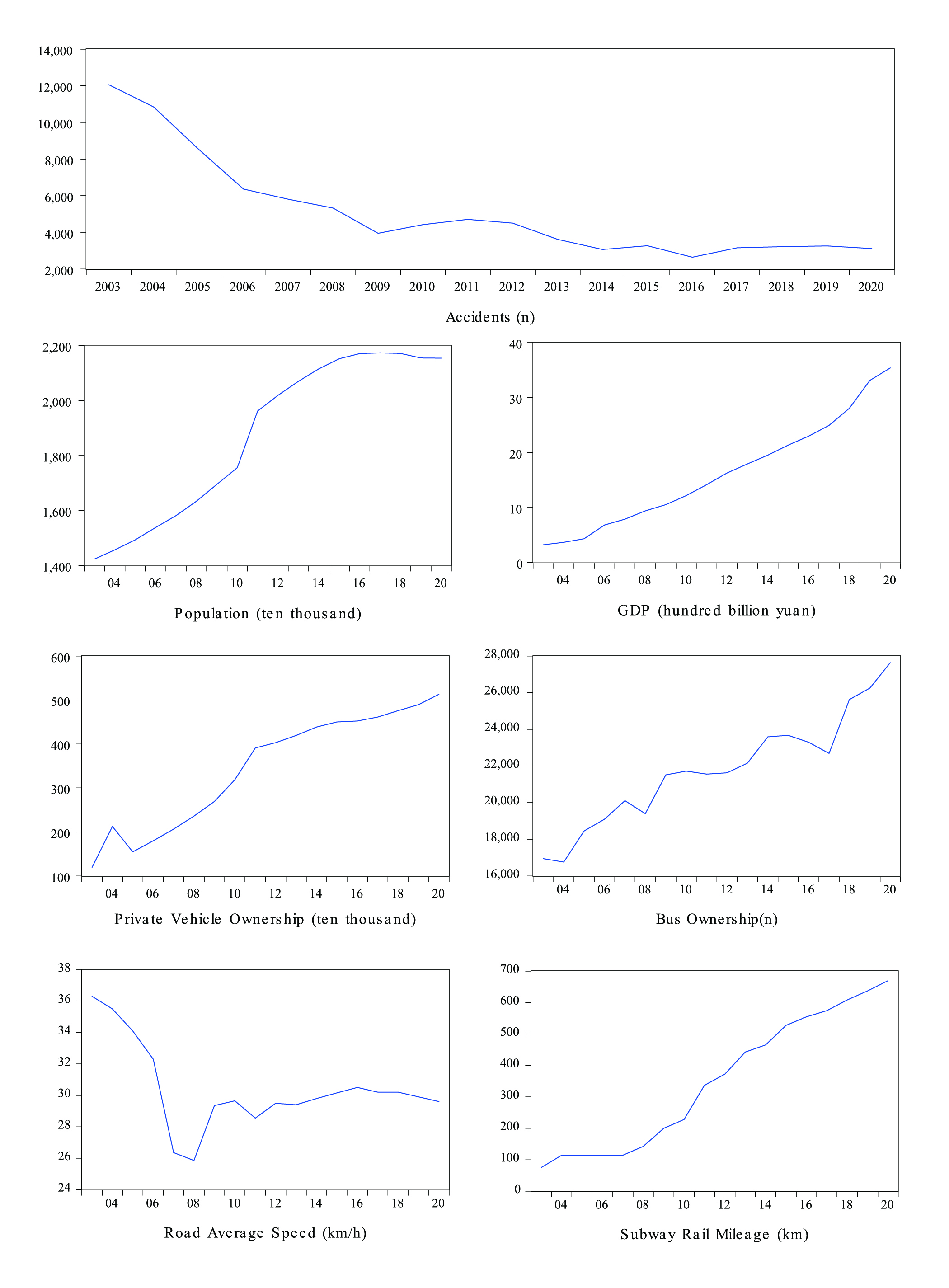}
	\caption{Variables time-series distribution}
	\label{FIG:1}
\end{figure}

Figure \ref{FIG:1} reveals the evolutions of these time series. As can be seen in Figure \ref{FIG:1}, the number of traffic accidents in Beijing had declined significantly over time, from over 12,000 in 2003 to around 3,000 in 2020. 
Population and GDP were both on a growth trend, the population growth slowed around 2015, while GDP has been growing at a steady rate.
Private vehicle and bus ownership both tended to increase over time amidst fluctuations.
Subway rail mileage had been growing at a relatively steady rate, while the road average speed in 2020 was significantly lower compared to 2003.

\subsection{Data normalisation}

Tabel \ref{Tab:1} presents the descriptive statistical information of the dataset, as shown in the table, the dataset contained 18 yearly times series data, and the magnitudes of the variables were different from each other. To address the impact of different data magnitudes on the analysis results, we normalized all variables within the dataset to between 0 and 1 before being substituted into the model. Assume that the variable takes the value $X$ and the normalized variable takes the value $\bar{X}$ as follows:

\begin{equation}
    \bar{X}=\frac{X-X_{min}}{X_{max}-X_{min}}
\end{equation}

\begin{table}[width=.9\linewidth,cols=8,pos=h]
\caption{Summary of variables and descriptive statistics.}\label{Tab:1}
\begin{tabular*}{\tblwidth}{@{} LLLLLLLL@{} }
\toprule
Variables & Times Series & Observations & Mean & Median & Min & Max & SD\\
\midrule
Accidents() &  Yearly & 18 & 5101.44 & 4181.00 & 2639.00 & 12053.00 & 2743.98\\
Population(ten thousand) & Yearly & 18 & 1872.88 & 1989.90 & 1423.00 & 2172.90 & 292.01\\
GDP(billion yuan) & Yearly & 18 & 1618.19 & 1518.28 & 321.27 & 3537.13 & 994.11\\
Private Vehicle Ownership & \multirow{2}{*}{Yearly} & \multirow{2}{*}{18} & \multirow{2}{*}{343.83} & \multirow{2}{*}{396.85} & \multirow{2}{*}{119.50} & \multirow{2}{*}{513.00} & \multirow{2}{*}{130.88}\\
(ten thousand)&&&\\
Bus Ownership(ten thousand) & Yearly & 18 & 2.14 & 2.16 & 1.68 & 2.76 & 0.29\\
Subway Rail Mileage(km) & Yearly & 18 & 349.17 & 354.00 & 75.00 & 669.00 & 213.55\\
Road Average Speed(km/h) & Yearly & 18 & 30.40 & 29.85 & 25.85 & 36.30 & 2.70\\
\bottomrule
\end{tabular*}
\end{table}

\section{Methodology}

This paper developed a Bayesian vector autoregressive(BVAR) model to analyse urban macro factors of traffic accidents. The BVAR model is a special form of vector autoregressive(VAR) model. To understand the BVAR model, one can start with the VAR model.

\subsection{Vector Autoregressive model}

VAR models are multivariate time series analysis models which are used extensively for macroeconomic analysis since their introduction by \cite{sims1980macroeconomics}. They can  capture the stochastic trends which separate the long-run relations from the short-run dynamics of the generation process of a set of variables and describe the dynamic structure and the joint generation mechanism of the variables involved \citep{lutkepohl2013vector}. Consequently, VAR models can be applied to factor analysis of multivariate time series data. 
The $N\times1$ vector $Y=\left\{y_{1T},y_{2T},\ldots,y_{NT}\right\}\in\mathbb{R}^{N\times T}$ is a multivariate time series data. The number of observations is $N$ and the number of time series is $ T $. In the matrix $Y$, at any $t$-th time interval, the observed value is:

\begin{equation} 
y_{t}=\left(y_{1t},y_{2t},\ldots,y_{Nt}\right) \in\mathbb{R}^{N},
\end{equation}

Given multivariate time series data as $Y\in\mathbb{R}^{N\times T}$, the following linear expression for the vector autoregressive model exists for any $t$-th time interval:

\begin{equation} 
y_{t}=c+\sum_{k=1}^{d}A_{k}y_{t-k}+\epsilon_{t}, t=d+1,\ldots,T,
\end{equation}
where $c=(c_{1},\ldots,c_{N})^{T}$ is an $N\times 1$ vector of constants. $A_{k}\in\mathbb{R}^{N\times N},k=1,2,\ldots,d$ are the coefficient matrices of the vector autoregressive model. $\epsilon_{t}$ is a Gaussian noise vector satisfying $\epsilon_{t}\sim \mathcal{N}(0,\Sigma)$. $\Sigma$ is an $N\times N$ positive definite matrix. The traditional VAR model usually use the maximum likelihood method to estimate the parameters\citep{ni2005bayesian}.   

\subsection{Bayesian Vector Autoregressive model}

As we all know, the sample size plays an important role in the estimation of the parameters of this model because when the dimensionality of the time series sample is small, the estimates are not precise with the traditional method such as maximum likelihood and least squares method \citep{michail2020quantifying}.  In the case of a multivariate model there are usually a large number of parameters that would need to be estimated, it is difficult to obtain precise estimates for all the parameters applicable to a data set with a limited number of observations, without prior information or some form of parameter shrinkage. While traditional VAR models have a problem with the loss of degrees Bayesian techniques can be used to provide parameter estimates where the models include many variables and relatively little data.

In this paper, we used a Bayesian approach to estimate the var model to solve the problem of too many parameters and too few samples in multivariate model.
Unlike classical estimation methods, the basic idea of Bayesian estimation methods is to treat the parameters of the model to be estimated as random variables and obey a certain distribution, and then empirically give the prior distribution of the parameters to be estimated and combine it with the sample information. Bayes' theorem can be used to calculate the posterior distribution of the parameter to be estimated, resulting in an estimate with the estimated parameter\citep{ma2021application}. We assume that the VAR model does not contain a constant term, the model can be written in the following form:

\begin{equation}
    y_{it}=\sum_{j=1}^{N}\sum_{k=1}^{d}\alpha_{ijk}y_{jt-k}+\epsilon_{it}, i=1,\ldots,N, t=d+1,\ldots,T
\end{equation}
where $\alpha_{ijk}$ indicates the coefficient of the $k$-th order lag term $y_{j}$ of the variable $y_{j-k}$ in the $i$-th equation. If the random parameter $\alpha_{ijk}$ follows a normal distribution with mean $\delta_{ijk}$ and variance $S_{ijk}^{2}$, then the number of hyperparameters to be determined in the var model (3) is at least $2N^{2}d$, with $N^{2}d$ prior means and $N^{2}d$ prior variances.  Without considering the desirability of a priori information, it is quite complicated to reasonably assign the values of these  $2N^{2}d$ hyperparameters in general. Therefore, it is necessary to find ways to reduce the number of hyperparameters that need to be assigned, to determine reasonable values of hyperparameters, and to improve the predictive capability of the model. In setting the prior distributions, we follow standard practice and use the procedure developed in \cite{litterman1986forecasting} with modifications proposed by 
\cite{kadiyala1997numerical} and \cite{sims1998bayesian}. The Minnesota prior distribution is an effective method to solve this problem, and its basic assumptions include the following aspects:

\begin{enumerate}[(1)]
\itemsep=0pt

\item Normality: $\epsilon_{t} = \left( \epsilon_{1}, \epsilon_{2}, \cdots, \epsilon_{N} \right) \sim \mathcal{N}_{N} \left(0, \Sigma\right)$.

\item Independence: the covariance matrix $\Sigma$ and the model coefficient $\alpha_{ijk}$ are independent of each other.

\item The prior distribution of the covariance matrix $\Sigma$ is taken as the diffusion distribution:
$$\pi\left(\Sigma\right)\propto\vert\Sigma\vert^{-\left(N+1\right)/2},\sigma>0$$

\item The Model coefficients are mutually independent and obey normal distribution:
$$\alpha_{ijk}\sim \mathcal{N}\left(\delta_{ijk},S_{ijk}^{2}\right)$$

\item The mean value $\delta_{ijk}$ is determined according to the following formula:
$$\delta_{ijk}=\left\{\begin{array}{ll} 1,& i=j,k=1,\\ 
0,& \text{otherwise},
\end{array}\right.$$

\item The standard deviation $S_{ijk}$ can be decomposed as the product of four factors:
$$S_{ijk}=\gamma \cdot g\left(k\right)\cdot f\left(i,j\right)\cdot \frac{S_{i}}{S_{j}} $$
where $\gamma$ denotes the overall tightness, the magnitude of its value reflects the degree of confidence the analyst has in the prior information, and a smaller value of $\gamma$ represents a greater certainty of the prior information; $g\left(k\right)$ is the tightness of the $r$-th order lagged variable relative to the first-order variable, which indicates the reduction in the usefulness of past information over current information; function $f\left(i,j\right)$ is the tightness of the $j$-th variable in the $i$-th equation relative to the ith variable, and $S_{i}$ is the standard deviation of the univariate autoregressive model for variable $y_{i}$, and $S_{j}$ is the standard deviation of the univariate autoregressive model for variable $y_{j}$.

\end{enumerate}  

To facilitate the derivation, equation(3) can be written as a system of multivariate regressions:

\begin{equation} 
Y=X\beta+\epsilon, \quad \epsilon\sim \mathcal{N}\left(0,\Sigma\otimes I_{n}\right)
\end{equation}

The Minnesota  Priori shorthand for the model parameters is given as:

\begin{equation} 
    (\beta\mid\Sigma)\sim \mathcal{N}(\mu_{0},M_{0}),\quad \pi\left(\Sigma\right)\propto\vert\Sigma\vert^{-\left(N+1\right)/2}
\end{equation}
where $\mu_{0}$ is equal to 0 or 1, $M_{0}$ consists of $T$ diagonal matrices. According to the Bayesian theorem, under the Minnesota prior distribution condition, the joint posterior distribution density function of the parameter $(\beta,\Sigma)$ is:

\begin{equation} 
\begin{aligned}
 \pi\left(\beta,\Sigma \mid Y,X\right)&\propto\frac{1}{\vert\Sigma\vert^{\left(N+T+1\right)/2}}\exp \left\{-\frac{1}{2} \left[\left(Y-x\beta\right)^{T} \left(\Sigma\otimes I_{n}\right)^{-1}\left(Y-x\beta\right)+\left(\beta-\mu_{0}\right)^{T}M_{0}^{-1}\left(\beta-\mu_{0}\right)\right]\right\}\\
&\propto\frac{1}{\vert\Sigma\vert^{\left(N+T+1\right)/2}}\exp \left\{-\frac{1}{2} \left[\left(\beta-\beta_{B}\right)^{T}V^{-1}\left(\beta-\beta_{B}\right) -Y^{T}\left(\Sigma^{-1}\otimes I_{n}\right)^{-1}Y+\beta_{0}^{T}M_{0}^{-1}\beta_{0}\right]\right\}
\end{aligned}
\end{equation}
where 
$$\beta_{B}=V^{-1}\left[X^{T}\left(\Sigma\otimes I_n\right)^{-1}Y+M_{0}^{-1}\beta_{0}\right]$$
$$ V= \left[X^{T}\left(\Sigma\otimes I_n\right)X+M_{0}^{-1} \right]^{-1} $$

Obviously, for a given covariance matrix $\Sigma$, the conditional posterior distribution of $\beta$ is a multivariate normal-terminus distribution with mean $\beta_{B}$ and variance matrix $V$.

\begin{equation}
     (\beta\mid\Sigma;Y,X)\sim \mathcal{N}(\beta_{B},V)
\end{equation}

\section{Results and discussion}

The results of the BVAR model was estimated by Eviews 8.0 software based on the transport annual report data for Beijing from 2003 to 2020, and we built a VAR model as a comparison. 

\subsection{Model comparison results}

Whether VAR or BVAR model, the number of lags included needs to be selected. Based on the same data, the lag order selection criteria for the VAR model were the same as the BVAR model. As shown in Tabel \ref{Tab:2}, the optimal lag order was determined to order 1 in the VAR and BVAR models according to the five lag order evaluation indicators of LR\citep{reinsel1992vector}, FPE\citep{lutkepohl2005new}, AIC\citep{akaike1974new}, SIC\citep{neath1997regression} and HQIC\citep{hannan1979determination}.

\begin{table}[width=.9\linewidth,cols=7,pos=h]
\begin{threeparttable}
\caption{VAR and BVAR Lag Order Selection Criteria.}\label{Tab:2}
\begin{tabular*}{\tblwidth}{@{} LLLLLLL@{} }
\toprule
Log & LogL & LR & FEP & AIC & SIC & HQIC\\
\midrule
0 &  124.94 & NA & $2.22\times 10^{-15}$ & -13.88 & -13.53 & -13.84\\
1 & 246.48 & 128.68* & $7.68\times 10^{-19}$* & -22.41* & -19.66* & -22.14*\\
\bottomrule
\end{tabular*}
\begin{tablenotes}[flushleft]\footnotesize
    \renewcommand{\TPTtagStyle}[1]{\makebox[.6em][l]{#1}}
    \item[*] indicates lag order selected by the criterion. LR: sequential modified LR test statistic (each test at 5\% level). FPE: Final prediction error. AIC: Akaike information criterion. SIC: Schwarz information criterion. HQIC: Hannan-Quinn information criterion.
\end{tablenotes}
\end{threeparttable}
\end{table}

\begin{figure}[h!]
	\centering
	\begin{minipage}[t]{0.48\textwidth}
        \centering
		\includegraphics[scale=.75]{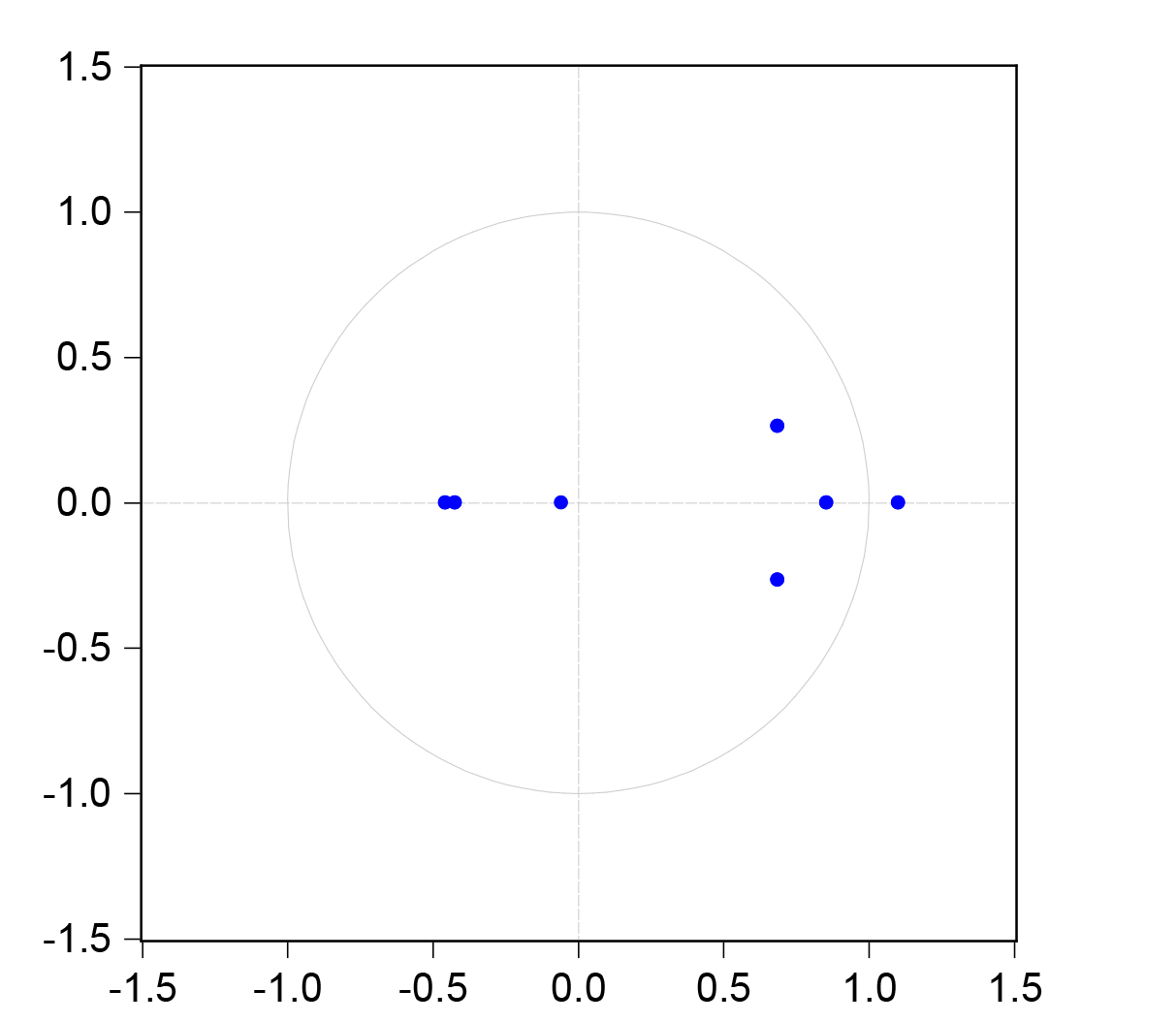}
	
	\subcaption{(a)Inverse Roots of VAR Characteristic Polynomial}
	\label{FIG2:(a)}
    \end{minipage}
	\begin{minipage}[t]{0.48\textwidth}
        \centering
		\includegraphics[scale=.75]{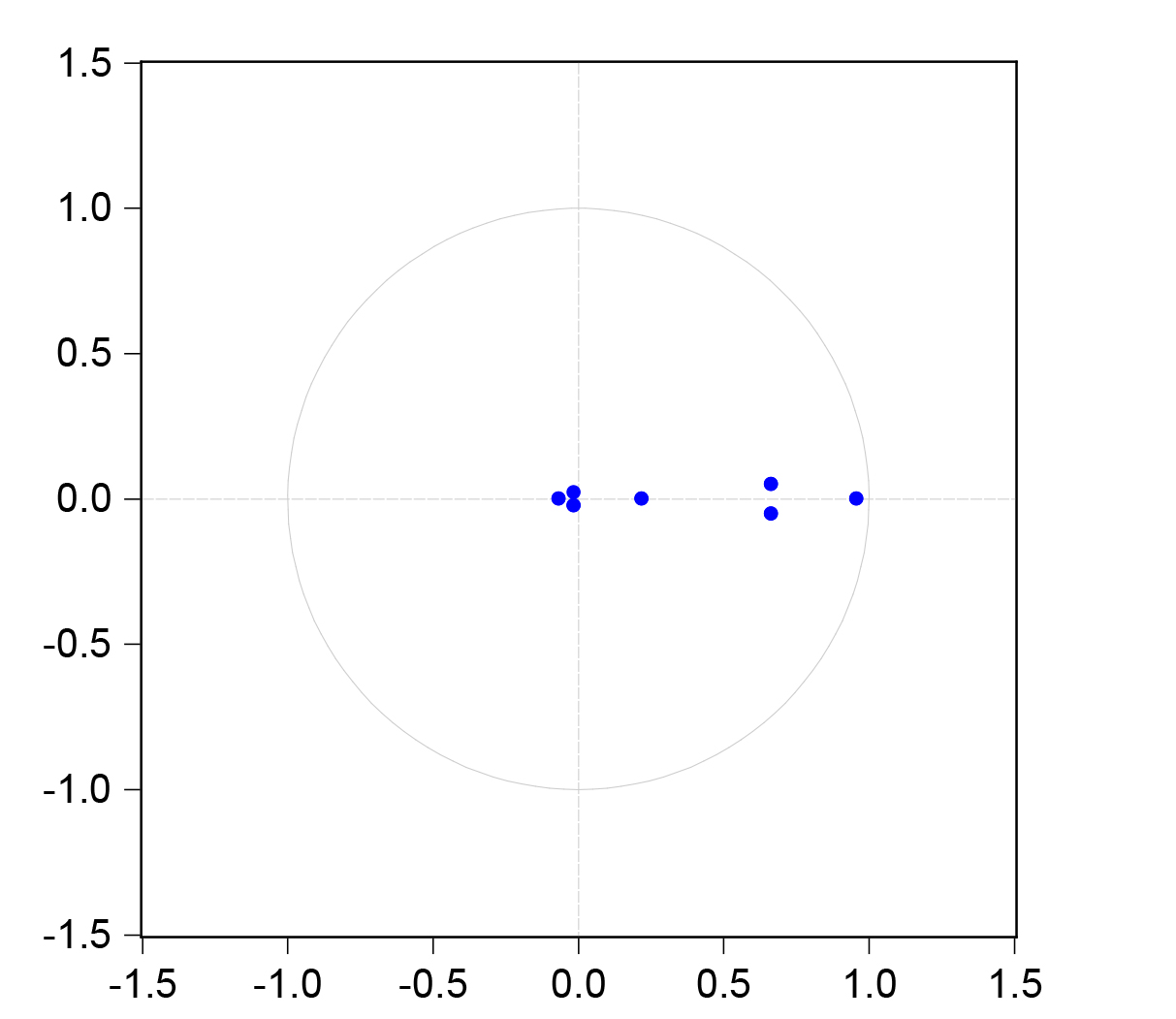}
	
	\subcaption{(b)Inverse Roots of BVAR Characteristic Polynomial}
	\label{FIG2:(b)}
    \end{minipage}
    \caption{Model Stability Setermination Criteria}
    \label{FIG:2}
\end{figure}

The stability test of the model is necessary for the credibility of the parameter estimates\citep{lutkepohl2013vector}. Before analysing the parameter estimation results, we determined the stability of the models. Figure \ref{FIG:2} reveals the stability test results of VAR and BVAR models. As shown in Figure (a), the stability test results of the VAR model had a root that was not within the unit circle, the results were unstable and the parameter estimation results cannot be used in the analysis results; as shown in Figure (b), the stability test results of the BVAR model were all within the unit circle, the parameter estimation results are plausible and can be used in the analysis results. Accordingly, in this study, the BVAR(1) model is more suitable to analysis the urban macroeconomic and transport factors of traffic accidents than VAR(1) model.

\subsection{Model estimation results and discussion}

\subsubsection{Overall estimation results}

Given the fact that 1 lag was used in the estimation of the BVAR, the equations can be presented in an easy-to-read format, as in Table \ref{Tab:2}, each column represents an autoregressive equation in one variable. As shown in Table \ref{Tab:2}, most of the autoregressive equations had R-squared values close to 1, and with standard errors close to 0, the model results were plausible. In overall terms, the parameter estimation results revealed that the number of traffic accidents were affected by all relevant indicators in this study. It has been noted that the coefficients in the parameter estimation results of BVAR are not indicative of the system behaviour, as such, impulse responses need to be examined to reach more concrete conclusions\citep{michail2020quantifying}.

\begin{table}[width=.9\linewidth,cols=8,pos=h]
\caption{BVAR model parameter estimation results}\label{Tab:3}
\begin{tabular*}{\tblwidth}{@{} LLLLLLLL@{} }
\toprule
  ~ & 1 & 2 & 3 & 4 & 5 & 6 & 7\\
\midrule
1. Accidents &  0.66 & -0.14 & 0.03 & 0.06 & -0.28 & 0.09 & 0.76\\
2. Population & 0.08 & 0.71 & -0.13 & 0.63 & -0.46 & 0.35 & 0.11\\
3. GDP & 0.11 & -0.12 & 0.67 & 0.28 & 0.79 & 0.36 & 0.45\\
4. Ownership of  Private Vehicle & -0.06 & 0.50 & -0.04 & 0.26 & 0.28 & 0.19 & 0.22\\
5. Ownership of bus & 0.15 & 0.14 & 0.21 & 0.34 & -0.02 & 0.10 & -0.04\\
6. Subway Rail & -0.27 & -0.30 & 0.37 & -0.48 & 0.15 & 0.17 & 0.53\\
7. Road Average Speed & 0.15 & 0.06 & -0.04 & 0.03 & -0.02 & 0.05 & 0.19\\
c & -0.03 & 0.08 & 0.04 & 0.02 & 0.34 & -0.11 & -0.12\\
\hline
R-squared & 0.95 & 0.99 & 0.99 & 0.97 & 0.95 & 0.99 & 0.74\\
S.E equation & 0.06 & 0.05 & 0.03 & 0.07 & 0.07 & 0.04 & 0.15\\
\bottomrule
\end{tabular*}
\end{table}

\begin{figure}[h!]
	\centering
	\includegraphics[scale=0.9]{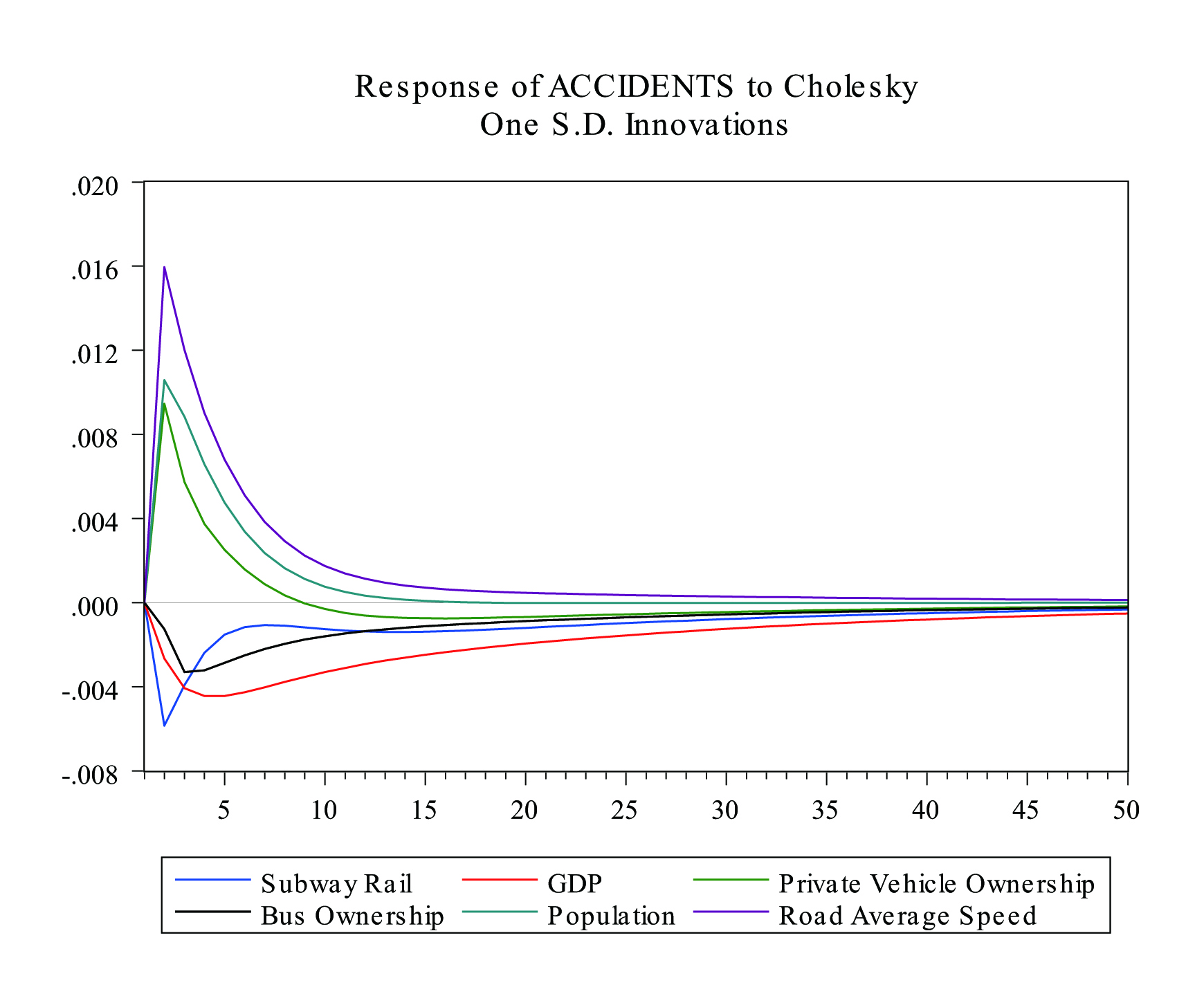}
	\caption{Impulse  Response of Traffic Accidents to Macro-level Variables}
	\label{FIG:3}
\end{figure}

Impulse response analysis is the tool which have been proposed for disentangling the relations between the variables in a VAR model\citep{lutkepohl2013vector}. From the study by \cite{li2019bayesian}, it can be inferred that a variable is considered to increase the number of traffic accidents if the majority of its cumulative impulse response is greater than zero, conversely if the majority of the cumulative impulse response is less than zero, the variable is considered to reduce the number of traffic accidents. Figure \ref{FIG:3} illustrates the impulse responses from the BVAR estimation results, a detailed analysis and discussion of the impulse response results shown in follows.

\subsubsection{Macroeconomic factors}

Concerning the population variable, the impulse response of accidents increased from 0 and reached a maximum value in period 2, then decreased rapidly to zero in period 10, and levelled off in period 13 and was greater than zero. These results suggest that population increase had the potential to increase the number of traffic accidents in the short term. A possible explanation for this might be that an increase in urban population would lead to an increase in travel demand, which could further increase the incidence of road traffic accidents. \cite{lee2008analysis} disclaimed that the rapid increase of travel demand may influence on the high rates of traffic accident.  These results are in accord with recent study in Zhongshan, China on  demographic characteristics and traffic accidents indicating that population has a significant positive influence on traffic accidents\citep{wu2020economic}.

In terms of the GDP variable, the impulse response of accident was also 0 in period 1, decreased to minimum in period 5, then increased and stabilized in period 30 and was consistently less than 0.  These results indicate that the increase in GDP could reduce the number of traffic accidents, and the effect would last for a long time. These results are contrary to the findings of a study in Algeria which indicated that that the number of traffic accidents in Algeria is positively influenced by the GDP per capita in the short and long term\citep{bougueroua2016economic}. This inconsistency may be due to that, in Algeria, GDP growth has led to an increase in private motor vehicles, which will increase the frequency of traffic accidents, whereas in Beijing, GDP growth has led to an improvement in public transport facilities and traffic safety facilities, which will reduce the risk of traffic accidents. Some studies also claimed that the increase of GDP has negative impacts on the number of fatalities in high-income countries, but it behaves oppositely in low-income countries\citep{van2000economic,bener2011road,yusuff2015impact}.

\subsubsection{Macro-transport factors}

Regarding the road average speed, we found that the impulse response of road traffic accidents was zero in the current period, then increased and reached a maximum in period 2, then began to decrease, and then stabilized and converged to 0 in period 20. In addition, this impulse response was mostly greater than zero in and greater than the other variables throughout the variation. These results show that an increase in the average speed of vehicles on urban roads might increase the frequency of urban accidents, and in the six variables of this model, average road speed had the greatest positive impact on traffic accidents, and the effects last for a long time. 
It seems possible that these results are due to that the increase in speed will reduce the driver's reaction time before an accident occurs, which will increase the risk of an accident occurring.
These are consistent with the findings of numerous studies analysing the correlation between vehicle speed and traffic accidents, with studies such as \cite{aljanahi1999speed}, \cite{elvik2013re} and \cite{sugiyanto2018determining} all pointed out that an increase in vehicle speed increases the risk of traffic accidents. Therefore, speed limits are one of the effective means to reduce urban traffic accidents.

We found the impulse response of accident to private vehicle ownership was zero in period 1, increased to a maximum in period 2, then decreased and stabilized to 0 in period 15. Furthermore, the majority of the impulse response was greater than zero and only lower than road average speed. 
From these results, it can be inferred that the increase in private vehicles ownership had a positive effect on urban traffic accidents, and the magnitude of this effect was second only to that of road average speed. 
Similar to population factors, a possible reason for these results is that an increase in private motor vehicles increases the demand for road traffic and, in turn, the incidence of accidents.
These results further support the idea of traffic safety impact analysis in Hong Kong that the growth of private vehicle ownership will increase the rate of traffic accidents\citep{li2018exploring}. Consequently, controlling the number of private motor vehicles plays an important role in reducing the number of urban traffic accidents.

With respect to subway rail mileage, the impulse response was also 0 in period 1, dropped to a minimum in period 2, then rose and stabilized in period 7, and converged to 0 approximately in period 50. This impulse response was the lowest response of six variables and mostly less than zero. These results provide important insights that the increase in subway rail mileage could reduce the number of traffic accidents, although the impact would diminish over time, their duration was long. Meanwhile the negative impact of subway rail mileage on traffic accidents was greater than other factors. 
These results are likely to be related to the traffic demand which shifts from roads to urban rail due to increased subway rail mileage.
This finding broadly supports the work of other studies in this area linking subway rail mileage with urban traffic accidents, for instance, \cite{litman2004rail,litman2005impacts} stated that increased rail transit will reduce traffic accidents rates, traffic injuries and fatalities, and \cite{mohapatra2015economic} also indicated that rail transit can reduce the urban traffic accidents.

About the bus ownership, the accidents impulse response decreased from 0 to minimum in period 3 and 4, then increased and stabilized in period 20 and was consistently less than 0. These results show that an increase in bus ownership was considered to reduce the number of traffic accidents, and similar to the subway rail mileage, although this negative effect diminished over this time, it persistd for a long period. 
It is difficult to explain this result, but the possible reason is that the road bus systems are more efficient and safer to transport than private transport systems, the improving of road bus system would shift the demand for private transport towards it, therefore may reduce the number of traffic accidents.
Similar findings are also evidenced by previous studies. For example, a report of \cite{duduta2015traffic} revealed that high quality bus systems can improve the traffic safety, 
\cite{tiwari2016impact} also pointed out that improving public transport infrastructure can enhance urban traffic safety.

\section{Conclusion and future directions}

This paper simultaneously explored the dynamic relationship between the macro-level economic and transport factors and urban traffic accidents. Based on a small sample size transport annual report data for Beijing, a BVAR model was developed to examine the contribution of six economic and transport variables, including GDP, population, private vehicle ownership, bus ownership, subway rail mileage and road average speed to the number of traffic accidents. 

The BVAR approach showed validities to explicitly explore the small sample size time series data structure in traffic safety field. The results confirmed that long-term or short-term relationships existed between the macro-level economic and transport factors and the number of urban traffic accidents. About macroeconomic factors, population growth might lead to an increase in the number of traffic accidents in the short term, while GDP growth had a negative long-term impact on the number of traffic accidents. In terms of macro-transport factors, urban metro rail mileage and bus ownership had a long-term negative effect on the number of traffic accidents, and subway rail mileage had the largest negative effect on the number of traffic accidents. Increases in average road speed and private motor vehicle ownership had the potential to increase traffic accidents, with average road speed had the largest positive effect on the number of traffic accidents over a longer period of time. The relevant government departments can reduce road traffic accidents by increasing economic investment in the construction of public transport facilities, limiting investment in private transport facilities and limiting road speed when formulating urban development direction and management policies. 

The analysis of the factors influencing traffic accidents in this paper has not yet considered the relationship between economic factors and transportation factors at the macro-level. In the future, the introduction of urban investment in transport facilities as an influencing factor can be considered to further analyse the influence process of macro-level economic and transport factors on traffic accidents.


\bibliographystyle{cas-model2-names}

\bibliography{cas-refs}


\end{document}